\documentclass[aps,prl,twocolumn,groupedaddress,floatfix,sort&compress,longnamesfirst]{revtex4-2}
\usepackage{amsmath,amssymb}
\usepackage{bm,bbm}
\usepackage{hyperref}
\usepackage{graphicx}

\hypersetup{
colorlinks=true,
linkcolor=blue,
filecolor=blue,
urlcolor=blue,
citecolor=blue,
}
\graphicspath{{figure/}}
\begin{document}

\title{Loss-induced Floquet non-Hermitian skin effect}
\author{Yaohua Li$^{1}$}
\author{Cuicui Lu$^{2,3}$}
\author{Shuang Zhang$^{3}$}
\author{Yong-Chun Liu$^{1,4}$}
\email{ycliu@tsinghua.edu.cn}
\affiliation{$^{1}$State Key Laboratory of Low-Dimensional Quantum Physics, Department of
Physics, Tsinghua University, Beijing 100084, China}
\affiliation{$^{2}$Key
Laboratory of Advanced Optoelectronic Quantum Architecture and Measurements
of Ministry of Education, Beijing Key Laboratory of Nanophotonics and
Ultrafine Optoelectronic Systems, School of Physics, Beijing Institute of
Technology, Beijing 100081, China}
\affiliation{$^{3}$Department of Physics,
University of Hong Kong, Hong Kong, China}
\affiliation{$^{4}$Frontier
Science Center for Quantum Information, Beijing 100084, China}
\date{\today }

\begin{abstract}
Non-Hermitian topological systems have attracted lots of interest due to
their unique topological properties when the non-Hermitian skin effect
(NHSE) appears. However, the experimental realization of NHSE conventionally requires non-reciprocal couplings, which are compatible with limited systems. Here we
propose a mechanism of loss-induced Floquet NHSE, where the loss provides
the basic source of non-Hermicity and the Floquet engineering brings about
the Floquet-induced complex next-nearest-neighbor couplings. We also extend
the generalized Brillouin zone theory to nonequilibrium systems to describe
the Floquet NHSE. Furthermore, we show that this mechanism can realize the
second-order NHSE when generalized to two-dimensional systems. Our proposal
can be realized in photonic lattices with helical waveguides and other
related systems, which opens the door for the study of topological phases in
Floquet non-Hermitian systems.
\end{abstract}

\maketitle







\textit{Introduction.}---Non-Hermitian systems exhibit rich topological
phases that are characterized by non-Hermitian topological invariants \cite%
{ashida_non-hermitian_2020,liu_second-order_2019,luo_higher-order_2019,kawabata_symmetry_2019,hu_non-hermitian_2021,wang_topological_2021}%
. One of their unique features is the appearance of the non-Hermitian skin
effect (NHSE) \cite%
{zhang_review_2022,lee_anomalous_2016,yao_non-hermitian_2018,song_non-hermitian_2019,borgnia_non-hermitian_2020,okuma_topological_2020,zhang_correspondence_2020,yi_non-hermitian_2020,guo_exact_2021,longhi_self-healing_2022,zhang_universal_2022}%
, which leads to the breakdown of conventional bulk-boundary correspondence
and the introduction of generalized Brillouin zone (GBZ) \cite%
{yao_edge_2018,yokomizo_non-bloch_2019,yang_non-hermitian_2020}, and shows
wide applications like chiral damping or amplification \cite%
{song_non-hermitian_2019_2} and anomalous lasing \cite{zhu_anomalous_2022}.
The NHSE has also been extended to higher-dimensional topological systems
with the emergence of the hybrid skin-topological effect, where the
topological edge states are further localized into the corners \cite%
{lee_hybrid_2019,li_gain-loss-induced_2022}, and it is a kind of
higher-order NHSE \cite{kawabata_higher-order_2020} originated from the
nontrivial interplay between the NHSE and the topological effect.

Recently, the NHSE has been observed in photonic systems \cite%
{weidemann_topological_2020}, acoustic systems \cite%
{zhang_acoustic_2021,zhang_observation_2021}, electrical circuits \cite%
{helbig_generalized_2020,zou_observation_2021} and through quantum dynamics
\cite{xiao_non-hermitian_2020}. However, all these realizations require
nonreciprocal couplings, which is not applicable in a large variety of
systems without nonreciprocity. At the same time, another mechanism through
the gain/loss is proposed, where the amplifying and dissipative behaviours
of the chiral current along different edges lead to the NHSE \cite%
{lee_anomalous_2016,yi_non-hermitian_2020,li_gain-loss-induced_2022}, but
the generation of the chiral edge current requires complex
next-nearest-neighbor couplings, which is difficult to implemented in
experiments.

On the other hand, topological phases have also been extended into
periodically driven systems, as known as Floquet topological insulators \cite%
{cayssol_floquet_2013,kitagawa_topological_2010,lindner_floquet_2011,jiang_majorana_2011,rudner_anomalous_2013,grushin_floquet_2014,fleury_floquet_2016,hu_dynamical_2020}%
. The Hamiltonians of those systems are periodic in time, offering the
opportunity to engineer the band structure in the quasienergy spectrum. The
Floquet topological phases have made great successes in the experiments \cite%
{flaschner_experimental_2016,peng_experimental_2016,maczewsky_observation_2017,mciver_light-induced_2020,wintersperger_realization_2020,zhang_superior_2021}%
, including the realization of the photonic topological phase \cite%
{rechtsman_photonic_2013} and the Haldane model \cite%
{jotzu_experimental_2014}. Recently, there are growing efforts in studying
the non-Hermitian topological phase transitions in nonequilibrium systems
\cite%
{zhou_non-hermitian_2018,hockendorf_non-hermitian_2019,zhou_dynamical_2019,zhang_non-hermitian_2020,weidemann_topological_2022,liu_symmetry_2022,zhu_hybrid_2022}%
. However, the relation between the Floquet driving and the NHSE remains
unknown.

In this paper, we uncover the mechanism of loss-induced Floquet NHSE, where
gain/loss combined with Floquet engineering leads to the NHSE. Here the
gain/loss provides the origin of non-Hermicity, while the Floquet
engineering plays a crucial role in generating the chiral current through
the Floquet-induced next-nearest-neighbor couplings. To describe the Floquet
NHSE, we also extend the generalized Brillouin zone theory to nonequilibrium
systems. Our proposal reveals a general mechanism to realize the NHSE, and we
show it can be extended to two dimensions to generate both the first-order
and second-order non-Hermitian skin effects. Importantly, our mechanism does not rely on the Floquet topology. Except for the second-order skin-topological effect, the first-order NHSE can be derectly generated by the Floquet-induced next-nearest-neighbor couplings even when the Floquet system remains trivial.

\begin{figure}[t]
\centering
\includegraphics[width=0.48\textwidth]{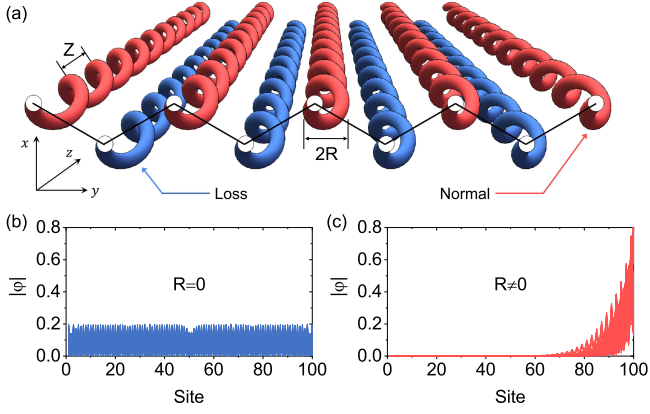}
\caption{(a) Sketch of the non-Hermitian helical waveguides. The waveguides
point to the $z$ direction and are arranged in a zigzag structure along the $%
y$ direction. There are two kinds of waveguides: the normal waveguides in
red and the lossy waveguides in blue. The helical radius is $R$, and the
pitch is $Z$. (b),(c) The profile of all the eigenstates of the effective
coupling matrix when $R=0$ (b) and $R=10$ $\mathrm{\protect\mu }$\textrm{m}
(c). There is no non-Hermitian skin effect for $R=0$ and there is the
non-Hermitian skin effect towards the right for $R\neq 0$. The distance
between two nearest-neighbor waveguides is $a=15$ $\mathrm{\protect\mu m}$.
Other parameters are $Z=1$ $\mathrm{cm}$, $k_{0}/2\protect\pi =2.4$ $\mathrm{%
\protect\mu m}^{-1}$, $c=1$ $\mathrm{cm}^{-1}$, $\protect\gamma =0.8$ $%
\mathrm{cm}^{-1}$.}
\label{fig:a}
\end{figure}

\textit{Floquet non-Hermitian skin effect.}---We first consider a Floquet
non-Hermitian photonic lattice, as illustrated in Fig. \ref{fig:a}(a). It is
a one-dimensional zigzag array of helical waveguides. The helical radius is $%
R$, and the pitch is $Z$. There are two waveguides in each unit cell, where
one of them is lossy and the other is normal, which is also similar to the
case of two waveguides with different losses. The loss in the waveguide can
be achieved, for example, by replacing the continuous waveguide with
periodic waveguide sections that have the same helical pattern. This
photonic lattice can be well described by a tight-binding model with
nearest-neighbor couplings, and the coupled equations can be written as \cite%
{rechtsman_photonic_2013}
\begin{equation}
i\partial _{z}\psi _{n}(z)=-i\gamma _{n}\psi _{n}(z)+\sum_{m=n\pm 1}ce^{i%
\bm{\mathrm{A}}(z)\cdot \bm{\mathrm{r}}_{mn}}\psi _{m}(z),  \label{eq:psi}
\end{equation}%
where $\psi _{n}(z)$ ($\gamma _{n}$) is the amplitude (loss) of the $n$th
waveguide, $c$ is the coupling strength, and $\bm{\mathrm{r}}_{mn}$ is the
displacement between waveguides $m$ and $n$. Here we assume $\gamma
_{n}=2\gamma $ for all the lossy waveguides, while the normal waveguides
have zero loss. After neglecting a global loss term $-i\gamma \psi _{n}$,
Eq. (\ref{eq:psi}) becomes the coupling equations of a non-Hermitian system
with balanced gain and loss. $\bm{\mathrm{A}}(z)=k_{0}R\Omega (-\cos \Omega
z,\sin \Omega z,0)$ is a $z$-dependent vector potential induced by the
helical shape of the waveguides, where $k_{0}$ is the wavenumber, and $%
\Omega =2\pi /Z$ is the frequency of the rotation.

We can rewrite the coupled equations as $i\partial_{z}\bm{\psi}(z)=H(z)%
\bm{\psi}(z)$, where $\bm{\psi}(z)=(\psi_{1},\psi_{2},\cdots)^{\mathrm{T}}$
is the amplitude vector, and $H(z)$ is the non-Hermitian coupling matrix.
The propagation of light along the waveguides can simulate the evolution of
the single-particle Schr\"{o}dinger equations, where the spatial dimension $%
z $ play the role of the time dimension and the coupling matrix is an analog
of the quantum Hamiltonian. The coupling matrix is $z$-dependent and has a
period $Z$: $H(z)=H(z+Z)$. So there are no static solutions for the coupling
equations. Instead, the solutions can be expressed in terms of the Floquet
states: $|\bm{\psi}(z)\rangle=e^{-i\varepsilon z}|\bm{\Phi}(z)\rangle$, with
$|\bm{\Phi}(z)\rangle=|\bm{\Phi}(z+Z)\rangle$, and $\varepsilon$ is the
quasienergy. The Floquet states are the eigenstates of the Floquet operator,
i.e. the evolution operator of a full period, $U(Z)=\mathcal{F}\mathrm{Exp}%
(-i\int_{0}^{Z}H(z)dz)$ ($\mathcal{F}$ denotes the spatial order along the $%
z $ axis). From the Floquet operator, we can define the effective coupling
matrix $H_{\mathrm{eff}}$ that satisfied $U(Z)=\mathrm{Exp}(-iH_{\mathrm{eff}%
}Z)$.

Figure \ref{fig:a}(b) and \ref{fig:a}(c) are the profiles of all the
eigenstates of the effective coupling matrix when $R=0$ and $R\neq 0$. In
the former case with straight waveguides ($R=0$), the system reduces to a
trivial chain with static nearest-neighbor couplings. The coupling matrix
satisfies $\mathcal{(PT)}^{-1}H\mathcal{PT}$ after neglecting the global
loss, where $\mathcal{P}$ is the mapping from the left side to the right
side, and $\mathcal{T}$ is the complex conjugation. In this case, the NHSE
is forbidden by the $\mathcal{PT}$ symmetry [Fig. \ref{fig:a}(b)]. Because
any eigenstate that is localized at one side will be mapped into the other
side after the $\mathcal{PT}$ operation. On the contrary, the helical
waveguides introduce a chiral potential encoded in the coupling phases and
break the $\mathcal{PT}$ symmetry. As shown in Fig. \ref{fig:a}(c), the
system with helical waveguides ($R\neq 0$) exhibits the NHSE with all the
eigenstates located at the right boundary. The NHSE is obtained through
Floquet-engineered couplings accompanied by non-Hermicity from gain/loss,
which does not require nonreciprocal couplings.

\begin{figure}[b]
\centering
\includegraphics[width=0.48\textwidth]{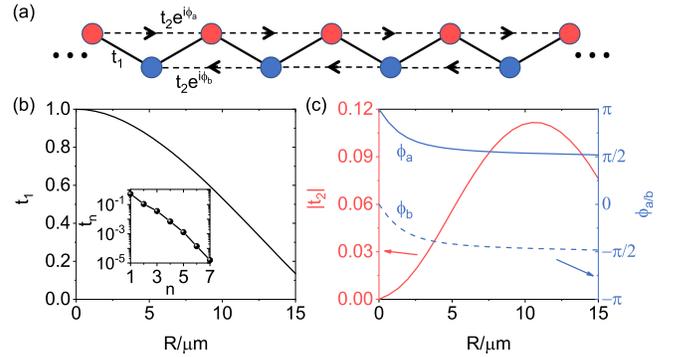}
\caption{(a) Effective coupling structure of the helical waveguides
considering the nearest-neighbor (solid lines) and next-nearest-neighbor
(dotted lines) couplings. The red (blue) circles denote the normal and lossy
waveguides, while the arrows indicate the phases of the
next-nearest-neighbor couplings. (b) The amplitude of the effective
nearest-neighbor coupling. The inset is the amplitudes of the long-range
couplings as a function of the distance $n$. (c) The amplitudes (red lines)
and phases (blue lines) of the effective next-nearest-neighbor couplings.
The coupling phases between normal (blue solid line) and loss (blue dotted
line) waveguides have opposite signs, corresponding to the opposite arrow
directions in (a). These results are obtained in the Hermitian case without
loss. Other parameters are the same as Fig. \protect\ref{fig:a}.}
\label{fig:b}
\end{figure}

\textit{Effective coupling structure.}---To reveal the physics of the
Floquet NHSE, we calculate the average strengths and phases of the couplings
in the effective coupling matrix $H_{\mathrm{eff}}$. Figure \ref{fig:b}(a)
is the sketch of the effective coupling structure, where $t_{n}$ denotes the
$n$th-nearest-neighbor couplings. As shown in the inset of Fig. \ref{fig:b}%
(b), the coupling strength is exponentially related to the distance $n$.
Consequently, we can neglect other long-range couplings except for the
leading two orders, i.e. the nearest-neighbor couplings $t_{1}$ [Fig. \ref%
{fig:b}(b)] and the next-nearest-neighbor couplings $t_{2}$ [Fig. \ref{fig:b}%
(c)]. In this case, the effective coupling equations can be approximately
written as
\begin{equation}
\begin{split}
i\partial _{z^{\prime }}\psi _{n}(z^{\prime })& \approx -i\gamma
_{n}^{\prime }\psi _{n}(z^{\prime })+t_{1}\psi _{n-1}(z^{\prime })+t_{1}\psi
_{n+1}(z^{\prime }) \\
& +t_{2}e^{i\phi _{2,n}}\psi _{n+2}(z^{\prime })+t_{2}e^{-i\phi _{2,n}}\psi
_{n-2}(z^{\prime }),
\end{split}%
\end{equation}%
%
%
%
where $\gamma _{n}^{\prime }$ is the effective loss, and $z^{\prime }=z/Z$
is the dimensionless coordinate. Here we have absorbed the coupling phases
of the nearest-neighbor couplings $t_{1}$ into the mode amplitudes $\psi
_{n}(z^{\prime })$. It is a coupling equation describing a static system
with the same coupling structure in the effective coupling matrix
(neglecting the long-range couplings for $n>2$). When $R\rightarrow 0$, the
effective nearest-neighbor coupling strength is $t_{1}=cZ$, corresponding to
the static photon tunneling strength between nearby waveguides. As shown in
Fig. \ref{fig:b}(b), the effective nearest-neighbor coupling strength
decreases when increasing the helical radius, due to the emergence of
long-range photon tunnelings. Among these long-range couplings, the
next-nearest-neighbor couplings play the most important role in the Floquet
no-Hermitian skin effect. Due to the chiral potential in the original
periodic coupling, the effective next-nearest-neighbor couplings between
normal and lossy waveguides, i.e. the upper and lower dotted line in Fig. %
\ref{fig:b}(a), possess coupling phases with opposite signs. Letting the
average values of two coupling phases being $\bar{\phi}_{2,n}=\phi _{a(b)}$
for $n\in \mathrm{odd}(\mathrm{even})$, we can obtain the two coupling
phases as shown in Fig. \ref{fig:b}(c). The blue solid (dotted) line denotes
the coupling phase between the normal (loss) waveguides. This asymmetry only
exists in the zigzag chain and will disappear when the waveguides are
arranged along a straight line. The asymmetric coupling phases can induce
the chiral edge current, which is the key requirement of the
gain-loss-induced NHSE \cite%
{lee_anomalous_2016,yi_non-hermitian_2020,li_gain-loss-induced_2022}. Since
the chiral edge current is independent on the loss, in the calculation of
effective coupling coefficients (Fig. \ref{fig:b}) we consider the Hermitian
case without loss. In a intuitive picture, the combination of loss and
multichannel interference with controllable phase from the Floquet coupling
effectively generates nonreciprocal couplings \cite{huang_loss-induced_2021},
which result in the NHSE.

\begin{figure}[t]
\centering
\includegraphics[width=0.48\textwidth]{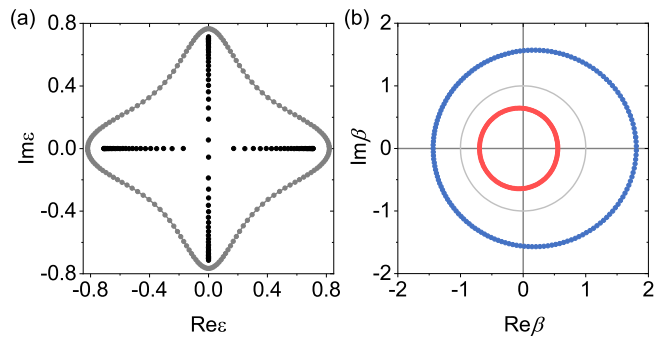} .
\caption{(a) The spectra of the photonic lattice with helical waveguides in
the periodic boundary condition (grey dots) and the open boundary condition
(black dots). (b) The generalized Brillouin zone of the system for $\protect%
\gamma =0.8$ $\mathrm{cm}^{-1}$ (blue dots) and $\protect\gamma =-0.8$ $%
\mathrm{cm}^{-1}$ (red dots). The grey line is a unit circle at the origin.
The helical radius is $R=10$ $\mathrm{\protect\mu m}$. Other parameters are
the same as Fig. \protect\ref{fig:a}}
\label{fig:c}
\end{figure}

\textit{Floquet generalized Brillouin zone.}---In the periodic boundary
condition along both $x$ and $y$ directions, the $k$-space coupling matrix
is given by $\mathcal{H}_{k}(z)=h_{1}(z)\sigma _{x}+h_{2}(z)\sigma
_{y}+i\gamma (\sigma _{z}-I)$, where $\sigma _{x,y,z}$ are the Pauli
matrices, $I$ is the identity matrix and the global loss term $-i\gamma I$
can be neglected as it does not affect the detailed band structure. The
coefficients $h_{1,2}$ are
\begin{gather}
h_{1}=c\cos \left[ \bm{\mathrm{A}}\cdot \bm{\mathrm{a}}_{1}\right] +c\cos %
\left[ \bm{\mathrm{A}}\cdot \bm{\mathrm{a}}_{2}-\sqrt{3}ka\right] , \\
h_{2}=-c\sin \left[ \bm{\mathrm{A}}\cdot \bm{\mathrm{a}}_{1}\right] -c\sin %
\left[ \bm{\mathrm{A}}\cdot \bm{\mathrm{a}}_{2}-\sqrt{3}ka\right] ,
\end{gather}%
where $k$ is the wave vector, $\bm{\mathrm{a}}_{1}=a(-\sin \pi /6,\cos \pi
/6)$, $\bm{\mathrm{a}}_{2}=a(-\sin \pi /6,-\cos \pi /6)$ and $a$ is the
distance between the nearest-neighbor waveguides. In Fig. \ref{fig:c}(a), we
plot the quasienergy spectrum in the periodic boundary condition (grey
dots), which forms a loop in the complex plane. The spectrum of a finite
lattice (black dots) lies in the interior of the loop. It indicates the
existence of the NHSE and the failure of the conventional band theory with
Bloch Hamiltonian \cite{zhang_correspondence_2020}.

\begin{figure*}[tbh]
\centering
\includegraphics[width=0.99\textwidth]{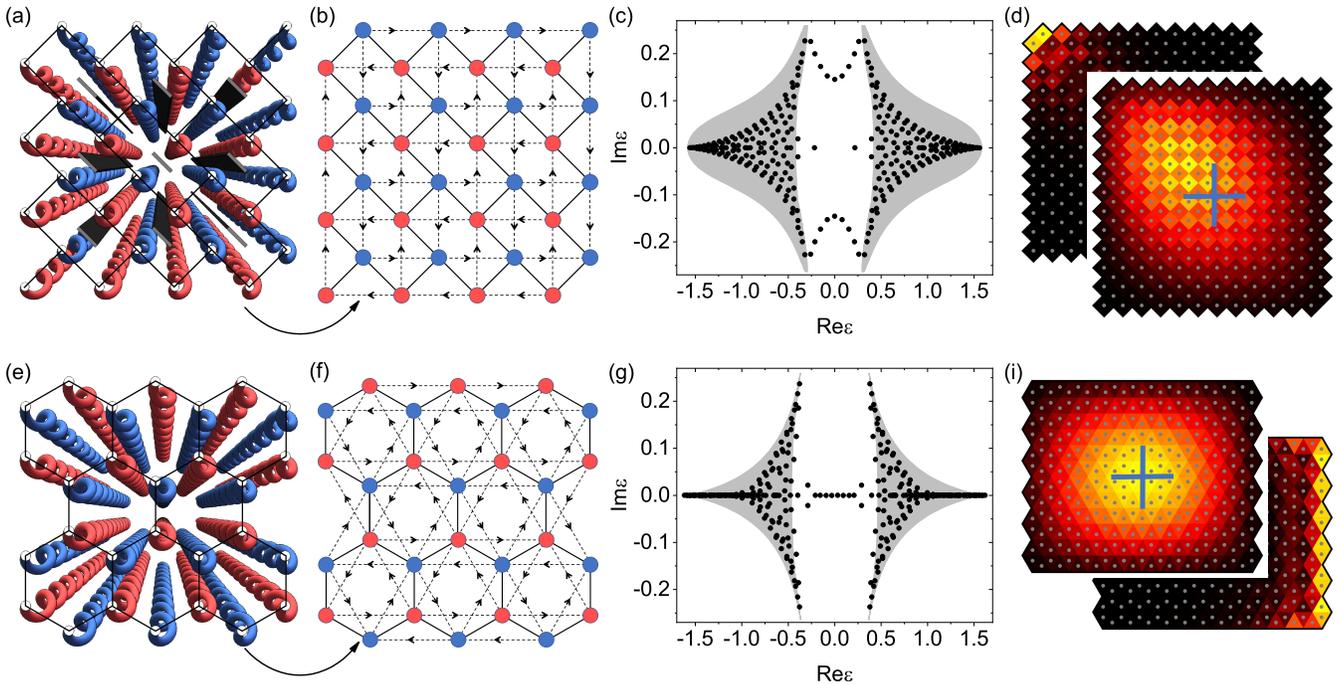}
\caption{Two-dimensional non-Hermitian skin effects in the two-dimensional
zigzag array (a)-(d) and the honeycomb array (e)-(i). (a),(e) Sketches of
the two structures. (b),(f) Effective coupling structures. (c),(g)
Quasienergy spectra in the periodic (grey areas) and open (black dots)
boundary conditions. (d),(i) Profiles of the bulk states (upper figures) and
the edge states (lower figures) normalized by the maximal value. The blue
crosses indicate the centers of the rectangles. The loss in the blue
waveguides is $\protect\gamma =0.3$ $\mathrm{cm}^{-1}$, and the helical
radius is $R=10$ $\mathrm{\protect\mu m}$. Other parameters are the same as
Fig. \protect\ref{fig:a}.}
\label{fig:d}
\end{figure*}

To recover the real-space spectrum, we extend the GBZ theory developed in
static systems to nonequilibrium systems. The basic idea of GBZ is the
replacement $e^{ik}\rightarrow \beta $ in the Bloch Hamiltonian to include
the exponential distribution of eigenstates due to the NHSE. The absolute
value of $\beta $, i.e., the GBZ can be unequal to unity, and it corresponds
to the exponential behavior of the eigenstates distributions. The
eigenstates will be localized rightwards (positive direction) for $|\beta
|>1 $ and leftwards (negative direction) for $|\beta |<1$. In static
two-band systems, the GBZ can be obtained by solving the characteristic
equation $f(\beta ,E)=\mathrm{det}[E-H(\beta )]=0$ under the condition that
the two specific solutions $\beta _{1,2}$ satisfy $|\beta _{1}|=|\beta _{2}|$%
. To obtain the GBZ in the nonequilibrium system, we need to solve the
characteristic equation of the effective coupling matrix: $f_{\mathrm{e}%
}(\beta ,E)=\mathrm{det}[E-H_{\mathrm{eff}}(\beta )]=0$, which, however, is
inaccessible as the effective coupling matrix can only be numerically
obtained. Instead, we calculate the GBZ by changing the absolute value of $%
\beta $ for a fixed argument and searching the eigenvalues with pure real or
imaginary values. In this case, we can obtain the GBZ that recovers the
real-space spectrum. Moreover, we check that all the GBZ obtained by this
method satisfy $f(\beta _{0},E_{0})=0=f(\beta _{0}^{\ast },E_{0})$. As shown
in Fig. \ref{fig:c}(b), when $\gamma >0$ (blue lines), the GBZ lies on the
outside of the unit circle, corresponding to the NHSE with rightwards
localization [see Fig. \ref{fig:a}(c)]. On the contrary, when $\gamma <0$
(red lines), the GBZ lies on the inside of the unit circle, corresponding to
the leftwards localization.

\textit{Two-dimensional generalizations.}---Here we show that the mechanism
of loss-induced Floquet NHSE can be generalized to two-dimensional systems.
Remarkably, rich phenomena including the first-order NHSE and the
second-order skin-topological effect can be realized. As shown in Fig. \ref%
{fig:d}(a), we first consider a two-dimensional array of helical waveguides,
which is a mesh of two zigzag models in both horizontal and vertical
directions. In order to take into account of the interplay between the
topology and the non-Hermicity, we consider that there are several plates
which prevent the couplings between special waveguides as plotted in Fig. %
\ref{fig:d}(a). The effective coupling structure considering the first two
orders of coupling is shown in Fig. \ref{fig:d}(b), where the arrows denote
the coupling phases. Without the losses, the effective static model in Fig. %
\ref{fig:d}(b) has a similar phase diagram of the Haldane model, with
topological phase transitions driven by the next-nearest-neighbor coupling
phases and on-site energies [see the Supplemental Material \cite{sup} for more details]. The Floquet-induced next-nearest-neighbor couplings can open
a topological band gap with the emergence of robust chiral edge states. In
the presence of losses, the spectra in the periodic boundary conditions
(grey area) and open boundary conditions (black dots) are very different, as
plotted in Fig. \ref{fig:d}(c), which reveals the existence of NHSE. As
shown in Fig. \ref{fig:d}(d), both the bulk states and the edge states
exhibit localization towards the corner, which represents the first-order
NHSE in two dimensions. A larger loss $\gamma $ leads to a clearer
localization behavior [see the Supplemental Material \cite{sup} for more details]. We note that the additional plates that block the couplings
between specific waveguides are essential for the introduction of the
topology, but the first-order NHSE is independent on the topology and still
exists without the additional plates.

In Fig. \ref{fig:d}(e), we further consider a honeycomb array of helical
waveguides. As shown in Fig. \ref{fig:d}(f), the Floquet-induced
next-nearest-neighbor couplings have coupling phases that exactly match
those in the Haldane model. Consequently, the honeycomb array can realize an
effective non-Hermitian Haldane model, and thus we can obtain the
second-order skin-topological effect \cite{li_gain-loss-induced_2022}. In
this case, only the topological edge states exhibit the NHSE, while the bulk
states keep extended with the same distribution as the Hermitian case [Fig. %
\ref{fig:d}(i)]. In the presence of the second-order skin-topological
effect, the conventional bulk-boundary correspondence is still valid, which
can be characterized by the non-Hermitian Chern number. We obtain the phase
diagram of the model and show that the skin-topological phase exists
approximately when the loss parameter $\gamma $ is smaller than the
effective nearest-neighbor coupling coefficient $t_{1}$ [see the
Supplemental Material \cite{sup} for more details]. %
%

\textit{Conclusion.}---In summary, we have proposed the mechanism of
loss-induced Floquet NHSE. The NHSE originates from the the combination of
gain/loss and the Floquet-induced next-nearest-neighbor couplings. We extend
the generalized Brillouin zone theory from static systems to nonequilibrium
systems and show it can solve the failure of conventional band theory with
Bloch Hamiltonians. Moreover, we show such a mechanism can also be used to
generate both the first-order and second-order NHSE in two dimensions based on different lattice structures. The first-order NHSE is similar with that in one dimension, which is independent from the system topology. As for the second-order skin-topological effect, we show the Floquet driving can perfectly reveal the phase diagram characterized by the non-Hermitian Chern numbers. Our
proposal paves the way for the investigation of non-Hermitian physics in
Floquet systems and gain/loss systems.

\begin{acknowledgments}
This work is supported by the Key-Area Research and Development Program of Guangdong Province (Grant
No. 2019B030330001), the National Natural Science Foundation of China (NSFC)
(Grant Nos. 12275145, 92050110, 91736106, 11674390, and 91836302), and the
National Key R\&D Program of China (Grants No. 2018YFA0306504).
\end{acknowledgments}


%

\end{document}